\documentclass[astrosymb]{aastex631}

\usepackage{comment}

\newcommand\eiso{E$_\mathrm{iso}$}
\newcommand\gb{GRB~221009A}

\newcommand\nTa{84}
\newcommand\nPo{26}
\newcommand\nTb{42}
\newcommand\nRe{33}
\newcommand\nga{184}
\newcommand\ngb{185}

\received{March 1, 2021}
\revised{April 1, 2021}
\accepted{January 22, 2025}

\submitjournal{APJ}

\shorttitle{GRB 221009A and Apparently Energetic GRBs}
\shortauthors{Atteia et al.}
\graphicspath{{./}{figures/}}

\begin{document}

\title{GRB~221009A and the Apparently Most Energetic Gamma-Ray Bursts}


\author[0000-0001-7346-5114]{Jean-Luc Atteia}
\affiliation{IRAP, Université de Toulouse, CNRS, CNES, UPS, Toulouse, France}
\author[0000-0000-0000-0000]{Laurent Bouchet}
\affiliation{IRAP, Université de Toulouse, CNRS, CNES, UPS, Toulouse, France}
\author[0009-0004-9396-5270]{Jean-Pascal Dezalay}
\affiliation{IRAP, Université de Toulouse, CNRS, CNES, UPS, Toulouse, France}
\author[0000-0003-3642-2267]{Francis Fortin}
\affiliation{IRAP, Université de Toulouse, CNRS, CNES, UPS, Toulouse, France}
\author[0000-0001-7635-9544]{Olivier Godet}
\affiliation{IRAP, Université de Toulouse, CNRS, CNES, UPS, Toulouse, France}
\author[0000-0000-0000-0000]{S\'ebastien Guillot}
\affiliation{IRAP, Université de Toulouse, CNRS, CNES, UPS, Toulouse, France}
\author[0000-0002-2652-0069]{Alain Klotz}
\affiliation{IRAP, Université de Toulouse, CNRS, CNES, UPS, Toulouse, France}

\author[0000-0000-0000-0000]{Fr\'ed\'eric Daigne}
\affiliation{UPMC-CNRS, UMR7095, Institut d'Astrophysique de Paris, F-75014, Paris, France}
\author[0000-0000-0000-0000]{Robert Mochkovitch}
\affiliation{UPMC-CNRS, UMR7095, Institut d'Astrophysique de Paris, F-75014, Paris, France}

\author[0000-0003-1835-1522]{Damien Turpin}
\affiliation{Université Paris-Saclay, Université Paris Cité, CEA, CNRS, AIM, Gif-sur-Yvette, France}



\begin{abstract}

Gamma-Ray Bursts (GRBs) are often referred to as the most luminous explosions in the Universe, due to their short and highly luminous prompt emission.
This apparent luminosity, however, does not reflect the true energy budget of the prompt emission, which is strongly beamed. 
Accurate estimations of the energy radiated during the prompt phase require taking into account the geometry of GRB jets, which remains poorly known. 
Nevertheless, one may establish the distribution of well measured quantities, like \eiso , the GRB isotropic equivalent energy, which encrypts crucial information about GRB jets, with the aim of providing constraints on the jets radiated energy. 
In this work, we study the bright end of the GRB isotropic equivalent energy distribution (hereafter called ``apparent energy''), using an updated sample of 185 apparently energetic GRBs with \eiso\ $\geq 10^{53}$ erg.
This new sample includes \gb , allowing to discuss this apparently super-energetic GRB in the context of the general \eiso\ distribution of long GRBs.
We describe the construction of the sample and compare two fits of the \eiso\ distribution with a simple power law and with a cutoff power law.
Our study confirms the existence of a cutoff around \eiso\ = $4\times10^{54}$~erg, even when \gb\ is included in the sample.
Based on this finding, we discuss the possible reasons behind the rapid decrease of the number of apparently energetic gamma-ray bursts beyond \eiso\ = $4\times10^{54}$~erg and the interpretation of \gb , the most apparently energetic GRB detected to date, in this context. 

\end{abstract}



\section{Introduction} 
\label{sec:intro}


Gamma-ray bursts (GRBs) appear as short and extremely luminous flashes of gamma-rays, with an isotropic equivalent bolometric luminosity that can reach or exceed $10^{53}~\mathrm{erg~s^{-1}}$ during the brief duration of the prompt emission. 
Integrating this luminosity 
over the duration of the prompt emission gives \eiso , the \textit{isotropic equivalent energy}, which can also reach extreme values, over $10^{54}~\mathrm{erg}$ \citep{Kulkarni1998, Cenko2010, Frederiks2013, Atteia2017}.

These large apparent energies are due to the particular geometry of GRBs whose prompt emission is produced in ultra-relativistic jets that are highly collimated \citep{Rhoads1997,Frail2001}.
Given this geometry, \eiso\ does not represent the total energy radiated in gamma-rays but the energy per solid angle emitted \textit{in our direction}.
Considering jets with typical opening angles of a few degrees \citep[e.g.,][]{Frail2001, Beniamini2019, Zhao2020} reduces the prompt energy output by a large factor, typically 100 to 1000, leading to energy budgets of few times $10^{51}~\mathrm{erg}$, comparable with the kinetic energy of typical core-collapse supernovae \citep[e.g.,][]{Woosley2006}. 

Despite its dependence on the observer's viewing angle, the study of the \eiso\ distribution presents some interest: firstly, \eiso\ is available for several hundreds of GRBs, allowing statistical studies and secondly, it conveys useful information on GRB jets, especially on its bright end, which is shaped by the most luminous jets seen on axis and little impacted by orientation effects.
In this article, we revisit the bright end of the long GRBs (LGRBs) isotropic energy distribution after the detection of GRB 221009A, whose isotropic energy exceeded all previous measurements. Our GRB sample is about two times larger than in our previous work \citep[][hereafter A17]{Atteia2017}. 
To stress the dependence of \eiso\ on the viewing angle, we use in the following the term of 'apparent energy' to qualify this quantity.

After constructing a sample of \ngb\ long GRBs with \eiso\ $\ge 10^{53}$ erg in Section \ref{sec:sample}, we confirm in Section \ref{sec:fit} the existence of a cutoff at few $\ge 10^{54}$ erg \citep[A17;][]{Tsvetkova2017, Lan2023}, and we demonstrate that the existence of \gb\ does not modify this conclusion. 
In the last section (Section \ref{sec:discussion}) we address the question of whether the exceptional \eiso\ of \gb\ makes it an exceptional burst, and we discuss the possible origin of the observed \eiso\ cutoff .


In all this paper, we use a flat $\Lambda$-CDM model, with the cosmological parameters measured by the Planck collaboration \citep{Planck2014}, $\mathrm{H_0 = 67.3~km~s^{-1}~Mpc^{-1}}$ and $\mathrm{\Omega_m = 0.315}$ to keep the consistency with the work of \cite{Tsvetkova2017}, \cite{Tsvetkova2021} and \cite{Poolakkil2021}. 
Unless otherwise specified, errors are given at the 1$\sigma$ level.

\section{Sample construction} 
\label{sec:sample}

Our sample includes \ngb\ GRBs with \eiso\ $\geq 10^{53}$ erg detected up to the end of 2023, including \gb . 
This is about 40\% of the 450 known GRBs with a redshift.
The choice of GRBs with \eiso\ $\geq 10^{53}$ erg is a compromise between two requirements. On one side, a dynamic range in \eiso\ covering more than 1 order of magnitude and a number of GRBs $> 100$ needed to statistically characterize the bright end of the energy distribution. On another side, the choice to have a nearly volume complete sample, even if this last requirement is not strictly necessary for this study.
In the following, we call 'apparently energetic GRBs', GRBs with \eiso\ $\geq 10^{53}$ erg and 'apparently very energetic GRBs', those with \eiso\ $\geq 10^{54}$ erg.

Since the computation of \eiso\ requires measuring the redshift and the spectrum of the prompt emission, our sample contains essentially GRBs detected by both the Neil Gehrels Swift X-ray  Observatory (hereafter \textit{Swift}, \citealt{Gehrels2004}) and Konus-Wind (hereafter Konus, \citealt{Aptekar1995}) or \textit{Swift} and the \textit{Fermi} Gamma-ray Burst Monitor (hereafter GBM, \citealt{Meegan2009}). 
It includes apparently energetic GRBs from the catalogs of \cite{Tsvetkova2017}, \cite{Tsvetkova2021} and \cite{Poolakkil2021}, in this order. 

We start with the catalogs of \cite{Tsvetkova2017} and \cite{Tsvetkova2021}, which are mutually exclusive and respectively contain \nTa\ and \nTb\ apparently energetic GRBs, and complete the sample with \nPo\ apparently energetic GRBs listed in \cite{Poolakkil2021}, which are not in the Konus catalogs. 
We use the measurements of \eiso\ provided in these catalogs without recalculating them, considering that the instrument teams are the best placed to compute this parameter from the data recorded by their instrument.
These three catalogs use the same cosmological parameters \citep{Planck2014}, allowing the direct comparison of \eiso .

Finally, the sample is completed with \nRe\ apparently energetic GRBs detected up to the end of 2023, which occurred after the publication of the previous catalogs, including the Brightest Of All Times (BOAT) \gb , whose prompt emission has been measured by several instruments \citep{An2023, Burns2023, Frederiks2023b, Ripa2023, Mitchell2024, Savchenko2024}.
This sample more than doubles the number of apparently energetic GRBs presented in A17, which included 75 GRBs with \eiso\ $\ge 10^{53}$ erg.

\begin{figure}
\plotone{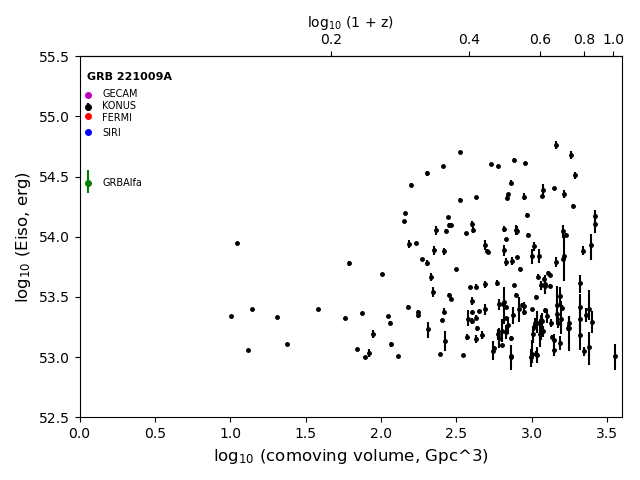}
\caption{Distribution of \ngb\ apparently energetic GRBs showing their isotropic equivalent energy \eiso\ and its uncertainty as a function of the volume enclosed within their distance.
The colored points show the position of \gb , based on the isotropic equivalent energy measured by GRBAlpha \citep[][green]{Ripa2023}, SIRI-2 \citep[][blue]{Mitchell2024},Fermi \citep[][red]{Lesage2022}, Konus \citep[][black]{Frederiks2023b}, GECAM \citep[][magenta]{An2023}. For a large fraction of the sample the error bars on \eiso\ are smaller than the symbol, including the SIRI, Fermi and GECAM measurements of \gb .
This figure emphasizes the peculiar location of \gb , which is both the closest of apparently energetic GRBs and the apparently most energetic of them.}
\label{fig:fig22}
\end{figure}

These bursts are listed in Table \ref{tab:tab1}, which gives their name, redshift and \eiso\ with its error.
A graphical view of this sample is shown in Figure \ref{fig:fig22}, which displays \eiso\ as a function of the comoving volume enclosed within the redshift of the GRB.
This plot emphasizes the very special position of \gb , which is at the same time the closest and the apparently most energetic burst in our sample, as emphasized by various authors \citep{An2023, Burns2023, Frederiks2023b, Lan2023, Laskar2023, Ripa2023, Tavani2023, Williams2023}.
Figure \ref{fig:fig22} additionally shows that our sample is not strongly affected by selection biases since apparently energetic GRBs, with \eiso\ $\sim 10^{53}$ erg, are detected up to the largest distances.

\section{Fitting the bright end of the \eiso\ distribution} 
\label{sec:fit}

In order to verify the existence of the energy cutoff discussed in A17, \cite{Tsvetkova2017} and \cite{Lan2023}, we fitted the differential
energy distribution above $10^{53}$~erg with a simple power law (PL), a cutoff power law (CPL) or a broken power law (BPL).
We fitted two samples: the \ngb\ GRBs in Table \ref{tab:tab1}, including the BOAT, and the same sample without the BOAT.
Regarding the BOAT, various authors have measured different values of its isotropic equivalent energy \eiso\ (see the references in Table \ref{tab:tab1}), and we have chosen the measurement of the Konus team (\eiso\ = $1.2\times10^{55}$~erg, \citealt{Frederiks2023b}) for consistency with the majority of the measurements in Table \ref{tab:tab1}, which also come from Konus. 
This is also an intermediate value between Fermi (\eiso\ = $1.0\times 10^{55}$~erg, \citealt{Lesage2023}) and GECAM (\eiso\ = $1.5\times 10^{55}$~erg, \citealt{An2023}). 
We have checked that this choice impacts only marginally the numbers in Table \ref{tab:tab2}  (at most a few \%).

For each function (PL, CPL, BPL) the best fit parameters are those which maximize the log-likelihood of the observed differential \eiso\ distribution. 
To determine the confidence intervals on the parameters, we use the classical approximation of the quantity $-2 \times \lbrack lnL(x_i,\theta) - lnL(x_i,\hat{\theta}) \rbrack $ by a chi-square distribution with $\nu$ degrees of freedom, where $\nu$ is the number of parameters of the fitting function. 
Here, $lnL(x_i,\hat{\theta})$ is the log-likelihood of the sample for the best fit parameters and $lnL(x_i,\theta)$ the log-likelihood of the same sample for an arbitrary value of $\theta$.
After a verification that the statistical uncertainties dominate over the contribution of the uncertainties on \eiso , we provide the best fit parameters and their errors computed with this procedure in Table \ref{tab:tab2}.

These fits can be used to test the existence of the cutoff.
A Kolmogorov-Smirnov (KS) test shows that the addition of a cutoff around $4\times 10^{54}$ erg increases the probability of the fits from about 1\% to about 50\%. 
The Bayesian Information Criterion Criterion (BIC, \citealt{Schwarz1978}) is often used to compare two fits of the same sample. The BIC is given by $\mathrm BIC = -2 ln(L) + k ln(N)$, where k is the number of fit parameters N the number of points in the sample and L the maximum likelihood of the sample. 
Adding a parameter improves the fit, increases the likelihood and usually reduces the BIC. 
The BIC reduction offers a way to measure the improvement brought by the supplementary parameter.
The addition of a parameter is usually considered as statistically justified if the BIC reduction exceeds 6. 
With respect to the PL fit, both the CPL and the BPL lead to significant reductions of the BIC (16 and 17 respectively), fully justifying the addition of the cutoff from a statistical perspective. 
The BIC, however, does not suggest a preferred model between CPL and BPL, from statistical grounds only.
Finally, we note that including the BOAT changes the values of the parameters slightly, without removing the need for a cutoff. 

\startlongtable
\begin{deluxetable*}{llcccccc}
\tablenum{2}
\tablecaption{Best fit parameters of the slope of the differential \eiso\ distribution, for the power law (PL), the cutoff power law (CPL) and the broken power law (BPL) models. The KS test probability and the BIC reduction measure the statistical justification of the cutoff (see text).}
\tablewidth{0pt}
\tablehead{
\colhead{Fit} & \colhead{Sample} & \colhead{Slope 1} & \colhead{E$_\mathrm{break}$} & \colhead{Slope 2} & \colhead{KS test} & \colhead{P(${\mathrm E_{\mathrm{iso}}}>1.2\times 10^{55}$)} &  \colhead{BIC} \\
 &  &  & \colhead{($\mathrm{10^{54}~erg}$)} & & probability & (p$_1$) & reduction \\
}
\startdata
\noalign{\smallskip}
PL  & 184 GRBs & $-1.71 \pm 0.05$ & -- & -- &  0.011  & $3.3\times 10^{-2}$ & N/A \\ 
 &  184 GRBs + BOAT & $-1.70 \pm 0.05$ & -- & -- & 0.013 & $3.4\times 10^{-2}$ & N/A \\ 
\noalign{\smallskip}\hline\noalign{\smallskip}
CPL & 184 GRBs &  $-1.26 \pm 0.18$ & $3.2_{-1.1}^{+2.6}$ & -- & 0.45 & $7.0\times 10^{-4}$ & 19.5 \\ 
 & 184 GRBs + BOAT & $-1.32 \pm 0.17$ & $4.2_{-1.5}^{+3.9}$ & -- & 0.55 & $1.7\times 10^{-3}$ & 16.3 \\ 
\noalign{\smallskip}\hline\noalign{\smallskip}
BPL & 184 GRBs &  $-1.48 \pm 0.13$ & $5.0_{-1.5}^{+1.4}$ & $\leq -3.6$ & 0.51 & $3.5\times 10^{-6}$ & 23.7 \\ 
 & 184 GRBs + BOAT & $-1.48 \pm 0.14$ & $4.0_{-2.0}^{+2.3}$ & $-4.0_{-4.0}^{+1.6}$ & 0.50 & $1.2\times 10^{-3}$ & 17.4 \\ 
\noalign{\smallskip}
\enddata
\label{tab:tab2}
\end{deluxetable*}

Table \ref{tab:tab2} also indicates, for each distribution, the probability to get a GRB as apparently energetic as \gb. 
For the best fit CPL distribution, this probability is of the order of 0.1\% 
Considering only the shape of the apparent energy distribution, it is thus not surprising to observe one GRB as apparently energetic as \gb\ in a sample of nearly 200 apparently energetic GRBs:
\gb\ appears as an extreme event, but not as a clear outlier to this distribution.
This conclusion, however, does not take into account the proximity of the BOAT, a point which is discussed in Section \ref{sub:discuboat}.

\begin{figure}
\plotone{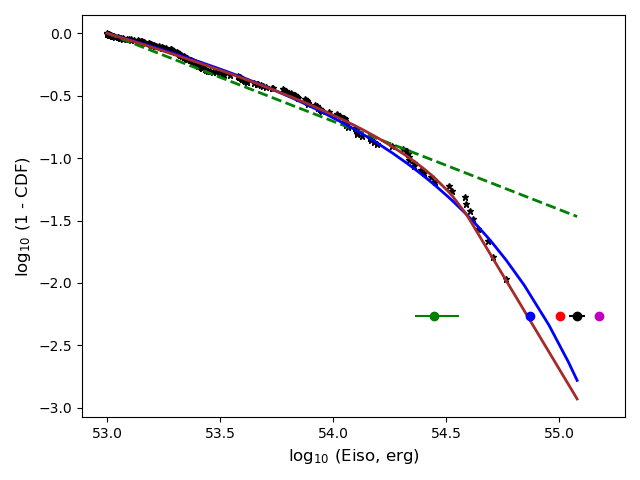}
\caption{Complementary cumulative energy distribution (CCDF = 1 - CDF) of \ngb\ GRBs, including \gb . 
The green, blue, red, black and magenta points show the isotropic equivalent energy of \gb\ respectively measured by GRBAlpha \citep{Ripa2023}, SIRI-2 \citep{Mitchell2024}, Fermi \citep{Lesage2022}, Konus \citep{Frederiks2023b} and GECAM \citep{An2023}. 
The green dashed line, the blue and brown solid line respectively show the best PL, CPL and BPL fits to the \eiso\ differential distribution.}
\label{fig:fig31}
\end{figure}

The cumulative best fit PL, CPL and BPL functions are shown in Figure \ref{fig:fig31}, on top of the observed cumulative \eiso\ distribution (note that 
the slope of the cumulative distribution is obtained by adding 1 to the numbers in Table \ref{tab:tab2}).
This figure clearly shows the fast decrease of the number of apparently energetic GRBs beyond a few times $10^{54}$~erg.

\subsection{Comparison with the work of \cite{Lan2023}} 
\label{sub:lan23}

In a similar work, \cite{Lan2023} studied the energy distribution of 355 GRBs with \eiso\ $\ge 10^{52}$ erg (319 long and 36 short), and compared \gb\ with apparently very energetic GRBs with \eiso\ $\ge 10^{54}$ erg. 
The number of GRBs with \eiso\ $\ge 10^{53}$ erg in their study (158 GRBs) is comparable to this work (185 GRBs).

\begin{figure}
\plotone{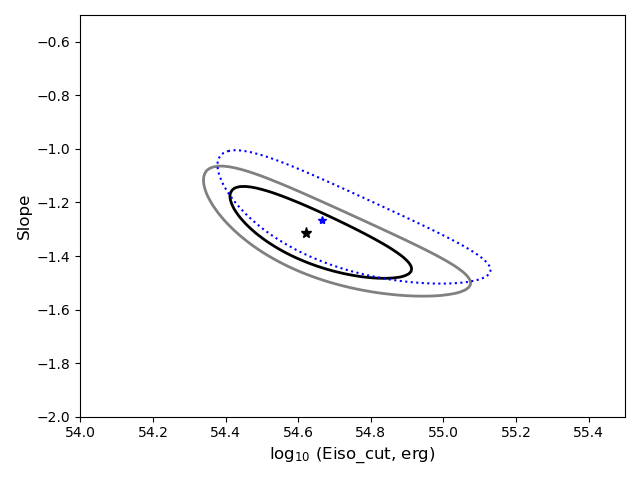}
\caption{Best fit parameters for the cutoff power law (CPL) in a plane showing the slope vs cutoff energy. The black star shows the best fit parameters for the sample in this study (with \gb\ included in the sample), and the solid contours show the 68 and 90\% confidence regions on these parameters. The blue star and dotted blue contour are the best fit and 90\% contour for the sample of \cite{Lan2023} restricted to GRBs with \eiso\ $\ge 10^{53}$ erg. }
\label{fig:fig33}
\end{figure}

They too conclude that a cutoff power law or a broken power law are strongly preferred over a simple power law.
It is not possible to directly compare the parameters in their study with ours, because they fit the cumulative energy distribution of GRBs with \eiso\ $\ge 10^{52}$ erg, while we work with the unbinned differential energy distribution of GRBs with \eiso\ $\ge 10^{53}$ erg. 
In order to perform a relevant comparison, we have (i) used their sample to calculate the best fit parameters for a CPL function fitting the cumulative energy distribution of GRBs with \eiso\ $\ge 10^{52}$ erg, and (ii) restricted their sample to GRBs with \eiso\ $\ge 10^{53}$ erg to calculate the best fit parameters for a CPL function fitting the differential energy distribution (see Figure \ref{fig:fig33} for this comparison). In both cases we find values compatible with theirs, showing the consistency of our analyses.

\section{Discussion} 
\label{sec:discussion}



\subsection{The case of GRB~221009A} 
\label{sub:discuboat}

Figure \ref{fig:fig22} clearly shows that \gb\ was a remarkable GRB, combining the shortest distance (by far) and the highest \eiso\ in our sample. 
We evaluate below the probability for \gb\ to be randomly drawn from of our sample.

What is the probability to observe a GRB closer than z = 0.151?
The sample contains 184 GRBs up to redshift z = 6.3 (excluding GRB~090429B at z = 9.38, which is the only burst in our sample detected beyond z = 6.3, where the identification of the afterglow requires near infrared observations).
Considering the density evolution proposed by \cite{Palmerio2022}, GRBs closer than z = 0.151 represent a fraction $\mathrm p_2 = 4.3\times10^{-4}$ of the number of GRBs within z = 6.3.
The chance to have 1 GRB closer than z = 0.151 among 184 events in our sample is thus $\mathrm p = 1 - (1 - p_2)^{184} \approx\ 0.08$.

This calculation does not take into account, however, the fact that \gb\ has also the highest \eiso . 
If one assumes the absence of luminosity or energy evolution with redshift, the probability of \gb\ in our sample is simply given by: $\mathrm p = 1 - (1 - p_1 \times p_2)^{184} = 1.4\times10^{-4}$ (where p$_1 = 1.7\times10^{-3}$ is given in Table \ref{tab:tab2}). 
While our ignorance of the joint \eiso\,-- redshift distribution prevents us from computing reliably the probability that the most apparently energetic GRB observed so far is also the closest, it is clear that the observation of a GRB as close and as apparently energetic as \gb\ is truly exceptional. This unique combination of distance and \eiso\ explains the outstanding brightness of  \gb . 
The rarity of \gb\ is confirmed by various authors, who evaluate the rate of such events from 1 per 10$^2$ to 10$^4$ yr \citep{Burns2023, Malesani2023, OConnor2023, Williams2023}.

Apart from its extreme isotropic energy, \gb\ seems to possess most characteristics of classical long GRBs, with respect to its prompt emission \citep{An2023, Burns2023, Frederiks2023b, Kann2023, Lesage2023, Ripa2023, Tavani2023, Savchenko2024}, to its afterglow \citep{Fulton2023, Laskar2023, Lhaaso2023, Williams2023} and to its association with a broad line supernova \citep{Fulton2023, Shrestha2023, Srinivasaragavan2023a, Blanchard2024, Kong2024}. 
The afterglow modelling indicates a narrow jet, with an opening angle in the range 0.6$^\circ$ to 2$^\circ$ \citep{An2023, Bright2023, Cao2023, Sato2023}. 
This narrow jet strongly reduces the energy budget of the jet, bringing it in the range $7\times10^{50}$ to $7\times10^{51}$ erg in gamma-rays, in agreement with classical GRBs \citep{Frail2001}.
Finally, the theoretical modelling also supports the interpretation of \gb\ as a classical long GRB with a structured jet with a very narrow core \citep{OConnor2023, Ren2023, Sato2023, Yang2023, Zhang2023}.
The true rate of events like \gb\ remains nevertheless difficult to estimate because the Earth is rarely within their beam and we have a single case.

This model however, does not explain the non-detection of GRBs as apparently energetic as \gb\ at higher redshifts.
According to the estimate above, there are about 2000 times more GRBs within z = 6.3 than within z = 0.151.
It is thus surprising that the only GRB with \eiso\ beyond $10^{55}$ erg was found so close to the Earth.
This has prompted some authors to investigate alternative ways to explain the BOAT. 
One such way is the gravitational lensing by an intervening stellar object. 
This possibility was studied by \cite{Bloom2022}, motivated by the low galactic latitude of \gb . His study eventually showed that the probability of lensing is very small, and significantly lower than the probability of observing a burst like \gb\ just by chance. 
Another approach has been taken by \cite{Finke2024}, who speculate on the existence of a population of nearby narrow-jet GRB population (BOAT-like GRBs), that are supposed to be present only below z $\approx 0.38$. Considering their high beaming and local origin, GRBs from this population would be detected at the rate of $\approx 3$ per century. 

The general consensus, however, is that \gb\ was a classical long and energetic GRB, whose occurrence close to us was a rare and lucky opportunity.
Under this assumption, we now discuss briefly the possible origin of the \eiso\ cutoff.

\subsection{Origin of the energy cutoff} 
\label{sub:cutoff}

Various authors \citep[A17;][]{Tsvetkova2017, Lan2023} have shown that apparently very energetic GRBs (close to the \eiso\ cutoff) do not seem to exhibit special properties of their prompt emission, like their duration or their position in the E$_\mathrm{peak}$--\eiso\ diagram (where E$_\mathrm{peak}$ is the peak energy of the spectral energy distribution measured at the source, \citealt{Amati2009}). 
While they often exhibit significant GeV emission, this emission seems to be the simple consequence of their large isotropic equivalent energy at keV -- MeV energies \citep{Ajello2019, SanchezRamirez2024}.
Overall, apparently very energetic GRBs appear as the high-energy end of the long GRB population, a conclusion that also applies to \gb , as discussed in \cite{An2023, Burns2023, Frederiks2023b, Kann2023, Lesage2023, Ripa2023, Tavani2023, Sears2024}.

In these conditions, we propose below a short generic discussion on the origin of the \eiso\ cutoff or break, within the framework of the collapsar model of GRBs.
We note that other explanations have been proposed in the context of other models, see for instance the work of \cite{Dado2022} for the cannonball model, but they are not addressed here.
The collapsar model involves a central engine (a black hole or a magnetar) created by the collapsing core of a massive star that is capable of launching transient ultra-relativistic jets sufficiently energetic to pierce the star's envelope without losing a significant fraction of their energy \citep[see, e.g.,][]{Woosley2006, Bromberg2011}.
One crucial ingredient of these models is the large angular momentum of the stellar core needed to avoid its direct collapse into the black hole and permit the creation of a temporary accretion disk \citep{MacFadyen1999}. 
The isotropic energy can be written as the product of $\mathrm E_{\gamma}$, the true energy released in gamma-rays by the jet emerging from the central engine, and the  beaming factor $\mathrm f_{\rm b}$. The beaming factor is the ratio of the flux emitted in our direction to the flux averaged over the whole sky. For a uniform jet, $\mathrm f_{\rm b} = {4\pi}/\Omega_{\rm j}$, where $\Omega_{\rm j}$ is the solid angle of the jet. 

\begin{equation}
\label{eq_eiso}
\mathrm E_{\rm iso} = \mathrm E_{\gamma} \times \mathrm f_{\rm b}
\end{equation}

The cutoff of the \eiso\ distribution requires a limitation on both quantities: the beaming factor and the energy released by the jet in gamma-rays, we briefly discuss these two quantities below.

While the isotropic energy covers a large interval of more than four orders of magnitude (from  $10^{51}$ to $10^{55}$ erg, for classical long GRBs), $\mathrm E_{\gamma}$ appears limited to a narrower interval of one or two orders of magnitude around $10^{50}$ erg \citep{Frail2001, Zhao2020}.
If the diversity in \eiso\ values is mostly the result of different jet opening angles, the 3 to 4 orders of magnitude ratio between $\mathrm E_{\rm iso}$ and $\mathrm E_\gamma$ corresponds to a jet opening angle of about 1 degree.
Indeed, the study of GRBs with the highest \eiso\ suggests that they have beaming factors of the order of $10^3$, corresponding to $\theta \approx 2.5^\circ$ and $\mathrm E_{\gamma} \geq 10^{51}$ erg (see for instance the discussion in section 4.2 of A17). 
Compared with these values, the beaming factor of \gb\ is still an order of magnitude larger, explaining its remarkable apparent energy with a rather typical $\mathrm E_\gamma$. 
The \eiso\ cutoff calls for an upper limit on $\mathrm f_{\rm b}$ ($\mathrm f_{\rm b} \le 10^3$), suggesting the existence of a minimum jet opening angle of about 1 degree.
We speculate that below this value the jet could simply not form or be subject to destructive instabilities during its propagation.

Turning now to $\mathrm E_{\gamma}$, it is the product of the jet energy $\mathrm E_{\rm j}$ by $\eta_{\rm j}$, the radiative efficiency of the jet.

\begin{equation}
\label{eq_egamma}
\mathrm E_{\gamma} =  {\mathrm E_{\rm j} \times \mathrm  \eta_{\rm j}}
\end{equation}

Assuming that GRBs close to the \eiso\ cutoff typically have $\mathrm f_{\rm b} \approx 1000$ and $\eta_{\rm j} \approx 0.2$ we can evaluate their jet power as $\mathrm E_{\rm j} \approx 2~10^{52}$~erg\footnote{We note an error on the exponent of E$_\mathrm{jet}$ in section 4.2 of A17, which should read 10$^{52}$ instead of 10$^{51}$, we apologize for this mistake.}. The observed cutoff would thus imply a limitation of the jet energy around this value.
In this context, it is interesting to note some theoretical limitations on the energy of relativistic jets produced by fast rotating magnetars or proto-magnetars  and those produced by stellar mass spinning black holes.
According to \cite{Metzger2011}, the maximum power of magnetar powered jets is limited to $\mathrm E_{\gamma} \sim 10^{51}$~erg.
Regarding stellar mass BH jets, the efficiency of the Blandford-Znajek mechanism \citep{BZ1977}, which is commonly invoked to extract the BH rotation energy and power GRB jets, strongly depends on the strength of the magnetic field and on the BH spin. 
It has  been shown recently that the angular momentum extracted by the BZ mechanism is very efficient to reduce the BH spin \citep{JacqueminIde2024}, leading to a limitation of the jet power, which cannot exceed 1 -- 2\% of the accretion energy, leading to $\mathrm E_{\rm j,max} \leq {\rm few}10^{52}$ \citep[e.g.][]{Wu2024}.
We speculate that the theoretical limitations on the power of the jets generated by the spinning magnetar model or the spinning BH model may explain the observed \eiso\ cutoff.

In any case, the final elucidation of the origin of the cutoff will require more observations of these rare apparently super-energetic events and detailed simulations of jet production and propagation, which are beyond the scope of this paper.

\section{Conclusions}
\label{sec:conclusion}

We have revisited the bright end of the GRB isotropic equivalent  energy distribution, after the detection of \gb . The main conclusions of our study are the following: 

\begin{itemize}

\item \gb\ appears as a classical long GRB, albeit with extreme apparent energy. This extreme apparent energy can be attributed to the geometry of its jet, which is among the narrowest detected so far. The lack of GRBs similar to \gb\ at larger resdhifts remains an open question, considering the small volume enclosed within z~=~0.151, compared to the volume explored with bright GRBs. With a single event, we cannot decide if \gb\ was a a highly unprobable, unique detection or if it belongs to a relatively nearby hypothetical population of narrow-jet GRBs, which makes the production of such GRBs more frequent at small redshifts.

\item The detection of \gb\ does not invalidate the existence of a cutoff around \eiso\ = $4\times10^{54}$ erg. The interpretation of this cut-off involves a limitation of the jet power combined with a dearth of ultra-narrow jets ($\theta_{\rm j} < 1^\circ$). While the first point is a generic prediction of GRB models, the second could be due to an intrinsic lack of ultra-narrow jets or their inability to successfully pierce the stellar enveloppe and still propagate over large distances without being destroyed by various instabilities. Detailed simulations of jet formation and propagation will be needed to test these two points.

\end{itemize}

From an observational perspective, it is also crucial to increase the currrent sample.
In this respect, the successful launch of \textit{SVOM} \citep{Wei2016, Atteia2022} on June $22^{\rm nd}$ 2024, should improve the situation in the coming years, with its capacity to provide both fast positions \textit{\`{a} la Swift} and broadband spectra of the prompt emission \textit{\`{a} la Fermi/GBM}.

\begin{acknowledgments}
The authors acknowledge the use of J. Greiner GRB page (\url{https://www.mpe.mpg.de/~jcg/grbgen.html}) to build the list of GRBs with \eiso\ $\geq 10^{53}$ erg used in this study. 
We thank the referee for comments that led us to clarify various points of the article.
\end{acknowledgments}

%

\vspace{5mm}
\facilities{Fermi(GBM), Konus-Wind, ESO-VLT}


\software{astropy \citep{Astropy2013, Astropy2018}
          }



\appendix

\section{Appendix information}


\startlongtable
\begin{deluxetable*}{lcccll}
\tablenum{1}
\tablecaption{\nga\ GRBs used in this study.}
\tablewidth{0pt}
\tablehead{
\colhead{GRB Name} & \colhead{Redshift} & \colhead{Eiso} & Satellite & References$^a$ \\
\colhead{} & \colhead{} & \colhead{($10^{52}$ erg)} &  & \\
}
\startdata
GRB 970828   & 0.96 & $ 26.2_{-  0.9}^{+  0.9}$ & KONUS & \cite{Tsvetkova2017} \\ 
GRB 971214   & 3.42 & $ 14.6_{-  0.6}^{+  0.6}$ & KONUS & \cite{Tsvetkova2017} \\ 
GRB 990123   & 1.60 & $213.3_{-  5.4}^{+  5.4}$ & KONUS & \cite{Tsvetkova2017} \\ 
GRB 990506   & 1.31 & $125.5_{-  4.3}^{+  4.3}$ & KONUS & \cite{Tsvetkova2017} \\ 
GRB 990510   & 1.62 & $ 17.4_{-  0.8}^{+  0.8}$ & KONUS & \cite{Tsvetkova2017} \\ 
GRB 990705   & 0.84 & $ 21.8_{-  0.8}^{+  0.8}$ & KONUS & \cite{Tsvetkova2017} \\ 
GRB 991208   & 0.71 & $ 23.3_{-  0.5}^{+  0.5}$ & KONUS & \cite{Tsvetkova2017} \\ 
GRB 991216   & 1.02 & $ 88.6_{-  1.1}^{+  1.1}$ & KONUS & \cite{Tsvetkova2017} \\ 
GRB 000131   & 4.50 & $181.7_{-  5.6}^{+  5.6}$ & KONUS & \cite{Tsvetkova2017} \\ 
GRB 000210   & 0.85 & $ 19.3_{-  0.5}^{+  0.5}$ & KONUS & \cite{Tsvetkova2017} \\ 
GRB 000911   & 1.06 & $ 64.9_{-  2.0}^{+  2.0}$ & KONUS & \cite{Tsvetkova2017} \\ 
GRB 000926   & 2.04 & $ 27.8_{-  1.1}^{+  1.1}$ & KONUS & \cite{Tsvetkova2017} \\ 
GRB 010222   & 1.48 & $107.2_{-  3.0}^{+  3.0}$ & KONUS & \cite{Tsvetkova2017} \\ 
GRB 020124   & 3.198 & $ 21.9_{-  2.8}^{+  3.1}$ & HET & \cite{Hjorth2003,Atteia2005} \\ 
GRB 020405   & 0.69 & $ 11.7_{-  0.2}^{+  0.2}$ & KONUS & \cite{Tsvetkova2017} \\ 
GRB 020813   & 1.25 & $ 75.8_{-  4.7}^{+  4.7}$ & KONUS & \cite{Tsvetkova2017} \\ 
GRB 030328   & 1.5216 & $ 37.9_{-  1.4}^{+  1.4}$ & HET & \cite{Maiorano2006,Sakamoto2005} \\ 
GRB 050401   & 2.90 & $ 46.4_{-  2.2}^{+  2.2}$ & KONUS & \cite{Tsvetkova2017} \\ 
GRB 050505   & 4.27 & $ 19.2_{-  2.2}^{+  2.7}$ & KONUS & \cite{Tsvetkova2021} \\ 
GRB 050730   & 3.9693 & $ 68.6_{- 25.5}^{+ 32.5}$ & KONUS & \cite{Tsvetkova2021} \\ 
GRB 050814   & 5.77 & $ 12.2_{-  3.6}^{+  4.0}$ & KW1 & \cite{Curran2008,Tsvetkova2021} \\ 
GRB 050820A  & 2.61 & $103.6_{-  3.6}^{+  3.6}$ & KONUS & \cite{Tsvetkova2017} \\ 
GRB 050904   & 6.295 & $147.7_{- 18.8}^{+ 20.4}$ & KONUS & \cite{Tsvetkova2021} \\ 
GRB 050922C  & 2.20 & $ 10.2_{-  2.4}^{+  2.4}$ & KONUS & \cite{Tsvetkova2017} \\ 
GRB 051008   & 2.77 & $ 83.4_{-  7.0}^{+  7.0}$ & KONUS & \cite{Tsvetkova2017} \\ 
GRB 051022   & 0.81 & $ 49.0_{-  1.0}^{+  1.0}$ & KONUS & \cite{Tsvetkova2017} \\ 
GRB 060124   & 2.30 & $ 32.8_{-  1.5}^{+  1.5}$ & KONUS & \cite{Tsvetkova2017} \\ 
GRB 060210   & 3.9122 & $112.7_{- 14.4}^{+ 13.4}$ & KONUS & \cite{Tsvetkova2021} \\ 
GRB 060418   & 1.4901 & $ 14.7_{-  0.9}^{+  0.9}$ & KONUS & \cite{Tsvetkova2021} \\ 
GRB 060607A  & 3.0749 & $ 16.3_{-  1.6}^{+  1.9}$ & KONUS & \cite{Tsvetkova2021} \\ 
GRB 060714   & 2.7108 & $ 10.5_{-  1.6}^{+  3.2}$ & KONUS & \cite{Tsvetkova2021} \\ 
GRB 060814   & 1.9200 & $ 41.4_{-  2.5}^{+  2.5}$ & KONUS & \cite{Tsvetkova2017} \\ 
GRB 060906   & 3.6856 & $ 19.3_{-  1.8}^{+  2.1}$ & KONUS & \cite{Tsvetkova2021} \\ 
GRB 060927   & 5.4636 & $ 22.6_{-  2.8}^{+  3.0}$ & KONUS & \cite{Tsvetkova2021} \\ 
GRB 061007   & 1.26 & $111.3_{-  4.0}^{+  4.0}$ & KONUS & \cite{Tsvetkova2017} \\ 
GRB 061121   & 1.31 & $ 30.4_{-  1.1}^{+  1.1}$ & KONUS & \cite{Tsvetkova2017} \\ 
GRB 061222A  & 2.09 & $ 26.0_{-  0.7}^{+  0.7}$ & KONUS & \cite{Tsvetkova2017} \\ 
GRB 070125   & 1.55 & $127.8_{-  6.4}^{+  6.4}$ & KONUS & \cite{Tsvetkova2017} \\ 
GRB 070328   & 2.06 & $ 77.7_{-  8.5}^{+  8.5}$ & KONUS & \cite{Tsvetkova2017} \\ 
GRB 070411   & 2.9538 & $ 15.6_{-  3.4}^{+  4.7}$ & KONUS & \cite{Tsvetkova2021} \\ 
GRB 070419B  & 1.9588 & $ 27.4_{-  2.8}^{+  3.0}$ & KONUS & \cite{Tsvetkova2021} \\ 
GRB 070521   & 2.0865 & $ 21.1_{-  0.7}^{+  0.6}$ & KW1 & \cite{Kruhler2015,Tsvetkova2017} \\ 
GRB 070721B  & 3.6298 & $ 27.0_{-  8.3}^{+ 11.7}$ & KONUS & \cite{Tsvetkova2021} \\ 
GRB 071003   & 1.60 & $ 38.5_{-  1.8}^{+  1.8}$ & KONUS & \cite{Tsvetkova2017} \\ 
GRB 071025   & 5.2 & $ 76.7_{-  6.1}^{+  7.5}$ & KONUS & \cite{Tsvetkova2021} \\ 
GRB 080207   & 2.0858 & $ 16.1_{-  1.8}^{+  2.4}$ & KONUS & \cite{Tsvetkova2021} \\ 
GRB 080319B  & 0.94 & $156.7_{-  1.9}^{+  1.9}$ & KONUS & \cite{Tsvetkova2017} \\ 
GRB 080319C  & 1.95 & $ 15.5_{-  1.6}^{+  1.6}$ & KONUS & \cite{Tsvetkova2017} \\ 
GRB 080411   & 1.03 & $ 23.9_{-  0.8}^{+  0.8}$ & KONUS & \cite{Tsvetkova2017} \\ 
GRB 080603B  & 2.69 & $ 10.0_{-  1.7}^{+  1.7}$ & KONUS & \cite{Tsvetkova2017} \\ 
GRB 080605   & 1.64 & $ 24.0_{-  0.5}^{+  0.5}$ & KONUS & \cite{Tsvetkova2017} \\ 
GRB 080607   & 3.04 & $217.1_{-  6.0}^{+  6.0}$ & KONUS & \cite{Tsvetkova2017} \\ 
GRB 080721   & 2.59 & $150.9_{-  6.3}^{+  6.3}$ & KONUS & \cite{Tsvetkova2017} \\ 
GRB 080804   & 2.2045 & $ 14.3_{-  0.6}^{+  0.6}$ & FERMI & \cite{Poolakkil2021} \\ 
GRB 080810   & 3.35 & $ 47.9_{-  1.6}^{+  1.6}$ & FERMI & \cite{Poolakkil2021} \\ 
GRB 080906   & 2.13 & $ 18.3_{-  2.8}^{+  3.0}$ & KONUS & \cite{Tsvetkova2021} \\ 
GRB 080916C  & 4.35 & $482.0_{- 39.0}^{+ 39.0}$ & KONUS & \cite{Tsvetkova2017} \\ 
GRB 081008   & 1.9685 & $ 14.4_{-  2.5}^{+  3.0}$ & KONUS & \cite{Tsvetkova2021} \\ 
GRB 081028A  & 3.038 & $ 20.0_{-  2.1}^{+  3.1}$ & KONUS & \cite{Tsvetkova2021} \\ 
GRB 081029   & 3.8479 & $ 20.5_{-  5.1}^{+  8.5}$ & KONUS & \cite{Tsvetkova2021} \\ 
GRB 081121   & 2.51 & $ 23.5_{-  1.0}^{+  1.0}$ & KONUS & \cite{Tsvetkova2017} \\ 
GRB 081203A  & 2.05 & $ 28.5_{-  9.6}^{+  9.6}$ & KONUS & \cite{Tsvetkova2017} \\ 
GRB 081221   & 2.26 & $ 39.7_{-  1.0}^{+  1.0}$ & KONUS & \cite{Tsvetkova2017} \\ 
GRB 081222   & 2.77 & $ 17.8_{-  1.2}^{+  1.2}$ & KONUS & \cite{Tsvetkova2017} \\ 
GRB 090102   & 1.55 & $ 20.0_{-  1.0}^{+  1.0}$ & KONUS & \cite{Tsvetkova2017} \\ 
GRB 090201   & 2.10 & $ 95.5_{-  2.8}^{+  2.8}$ & KONUS & \cite{Tsvetkova2017} \\ 
GRB 090323   & 3.60 & $581.0_{- 44.0}^{+ 44.0}$ & KONUS & \cite{Tsvetkova2017} \\ 
GRB 090328   & 0.74 & $ 10.9_{-  0.8}^{+  0.8}$ & KONUS & \cite{Tsvetkova2017} \\ 
GRB 090418A  & 1.608 & $ 14.2_{-  1.1}^{+  1.2}$ & KONUS & \cite{Tsvetkova2021} \\ 
GRB 090429B  & 9.38 & $ 10.2_{-  2.3}^{+  2.7}$ & KONUS & \cite{Tsvetkova2021} \\ 
GRB 090516   & 4.109 & $104.0_{-  4.0}^{+  4.0}$ & FERMI & \cite{Poolakkil2021} \\ 
GRB 090519   & 3.85 & $ 25.7_{-  1.6}^{+  1.6}$ & FERMI & \cite{Poolakkil2021} \\ 
GRB 090618   & 0.54 & $ 25.3_{-  0.5}^{+  0.5}$ & KONUS & \cite{Tsvetkova2017} \\ 
GRB 090715B  & 3.00 & $ 20.5_{-  1.9}^{+  1.9}$ & KONUS & \cite{Tsvetkova2017} \\ 
GRB 090812   & 2.45 & $ 27.2_{-  1.0}^{+  1.0}$ & KONUS & \cite{Tsvetkova2017} \\ 
GRB 090902B  & 1.822 & $402.0_{-  1.2}^{+  1.2}$ & FERMI & \cite{Poolakkil2021} \\ 
GRB 090926A  & 2.11 & $211.1_{-  5.3}^{+  5.3}$ & KONUS & \cite{Tsvetkova2017} \\ 
GRB 091003   & 0.8969 & $ 10.2_{-  0.2}^{+  0.2}$ & FERMI & \cite{Poolakkil2021} \\ 
GRB 091029   & 2.752 & $ 15.7_{-  2.5}^{+  2.4}$ & KONUS & \cite{Tsvetkova2021} \\ 
GRB 100414A  & 1.37 & $ 53.5_{-  1.1}^{+  1.1}$ & KONUS & \cite{Tsvetkova2017} \\ 
GRB 100606A  & 1.55 & $ 29.4_{-  2.3}^{+  2.3}$ & KONUS & \cite{Tsvetkova2017} \\ 
GRB 100724A  & 1.288 & $146.0_{-  0.8}^{+  0.8}$ & FERMI & \cite{Poolakkil2021} \\ 
GRB 100728A  & 1.57 & $114.0_{-  6.8}^{+  6.8}$ & KONUS & \cite{Tsvetkova2017} \\ 
GRB 100906A  & 1.73 & $ 24.9_{-  2.7}^{+  2.7}$ & KONUS & \cite{Tsvetkova2017} \\ 
GRB 110205A  & 2.22 & $ 63.0_{-  4.5}^{+  4.9}$ & KONUS & \cite{Tsvetkova2021} \\ 
GRB 110422A  & 1.77 & $ 74.7_{-  1.1}^{+  1.1}$ & KONUS & \cite{Tsvetkova2017} \\ 
GRB 110503A  & 1.61 & $ 21.3_{-  1.0}^{+  1.0}$ & KONUS & \cite{Tsvetkova2017} \\ 
GRB 110731A  & 2.83 & $ 31.5_{-  1.2}^{+  1.2}$ & KONUS & \cite{Tsvetkova2017} \\ 
GRB 110818A  & 3.36 & $ 19.2_{-  1.3}^{+  1.3}$ & FERMI & \cite{Poolakkil2021} \\ 
GRB 110918A  & 0.98 & $270.5_{-  9.9}^{+  9.9}$ & KONUS & \cite{Tsvetkova2017} \\ 
GRB 111008A  & 5.00 & $ 41.4_{-  7.1}^{+  7.1}$ & KONUS & \cite{Tsvetkova2017} \\ 
GRB 111123A  & 3.1516 & $ 39.0_{-  3.8}^{+  4.7}$ & KONUS & \cite{Tsvetkova2021} \\ 
GRB 120119A  & 1.73 & $ 40.2_{-  2.8}^{+  2.8}$ & KONUS & \cite{Tsvetkova2017} \\ 
GRB 120327A  & 2.813 & $ 19.1_{-  1.6}^{+  1.9}$ & KONUS & \cite{Tsvetkova2021} \\ 
GRB 120404A  & 2.876 & $ 10.3_{-  1.4}^{+  1.8}$ & KONUS & \cite{Tsvetkova2021} \\ 
GRB 120521C  & 6. & $ 19.5_{-  3.6}^{+  4.9}$ & KONUS & \cite{Tsvetkova2021} \\ 
GRB 120624B  & 2.20 & $282.0_{- 12.0}^{+ 12.0}$ & KONUS & \cite{Tsvetkova2017} \\ 
GRB 120711A  & 1.41 & $203.7_{-  4.6}^{+  4.6}$ & KONUS & \cite{Tsvetkova2017} \\ 
GRB 120712A  & 4.1745 & $ 17.5_{-  1.0}^{+  1.0}$ & FERMI & \cite{Poolakkil2021} \\ 
GRB 120716A  & 2.49 & $ 26.4_{-  2.5}^{+  2.5}$ & KONUS & \cite{Tsvetkova2017} \\ 
GRB 120802A  & 3.796 & $ 13.1_{-  1.7}^{+  1.7}$ & KONUS & \cite{Tsvetkova2021} \\ 
GRB 120909A  & 3.93 & $ 65.3_{-  2.6}^{+  2.6}$ & FERMI & \cite{Poolakkil2021} \\ 
GRB 120922A  & 3.1 & $ 44.6_{-  1.8}^{+  1.8}$ & FERMI & \cite{Poolakkil2021} \\ 
GRB 121128A  & 2.20 & $ 10.1_{-  0.4}^{+  0.4}$ & KONUS & \cite{Tsvetkova2017} \\ 
GRB 130408A  & 3.76 & $ 32.4_{-  6.1}^{+  6.1}$ & KONUS & \cite{Tsvetkova2017} \\ 
GRB 130427A  & 0.34 & $ 89.0_{-  0.5}^{+  0.5}$ & KONUS & \cite{Tsvetkova2017} \\ 
GRB 130505A  & 2.27 & $438.0_{- 10.0}^{+ 10.0}$ & KONUS & \cite{Tsvetkova2017} \\ 
GRB 130514A  & 3.6 & $ 61.5_{-  5.1}^{+  6.1}$ & KONUS & \cite{Tsvetkova2021} \\ 
GRB 130518A  & 2.49 & $216.0_{- 14.0}^{+ 14.0}$ & KONUS & \cite{Tsvetkova2017} \\ 
GRB 130606A  & 5.91 & $ 85.1_{- 21.1}^{+ 20.7}$ & KONUS & \cite{Tsvetkova2021} \\ 
GRB 130907A  & 1.24 & $384.0_{- 13.0}^{+ 13.0}$ & KONUS & \cite{Tsvetkova2017} \\ 
GRB 131011A  & 1.874 & $ 11.8_{-  0.5}^{+  0.5}$ & FERMI & \cite{Poolakkil2021} \\ 
GRB 131030A  & 1.29 & $ 32.7_{-  1.3}^{+  1.3}$ & KONUS & \cite{Tsvetkova2017} \\ 
GRB 131105A  & 1.69 & $ 15.3_{-  1.2}^{+  1.2}$ & KONUS & \cite{Tsvetkova2017} \\ 
GRB 131108A  & 2.40 & $ 54.0_{-  2.4}^{+  2.4}$ & KONUS & \cite{Tsvetkova2017} \\ 
GRB 131231A  & 0.64 & $ 21.1_{-  0.6}^{+  0.6}$ & KONUS & \cite{Tsvetkova2017} \\ 
GRB 140114A  & 3.0 & $ 16.0_{-  2.1}^{+  2.7}$ & KONUS & \cite{Tsvetkova2021} \\ 
GRB 140206A  & 2.73 & $ 25.0_{-  0.5}^{+  0.5}$ & FERMI & \cite{Poolakkil2021} \\ 
GRB 140213A  & 1.2076 & $ 10.6_{-  0.1}^{+  0.1}$ & FERMI & \cite{Poolakkil2021} \\ 
GRB 140304A  & 5.283 & $ 11.2_{-  0.9}^{+  0.9}$ & FERMI & \cite{Poolakkil2021} \\ 
GRB 140311A  & 4.954 & $ 25.9_{-  5.6}^{+  7.7}$ & KONUS & \cite{Tsvetkova2021} \\ 
GRB 140419A  & 3.96 & $228.0_{- 18.0}^{+ 18.0}$ & KONUS & \cite{Tsvetkova2017} \\ 
GRB 140423A  & 3.26 & $ 50.3_{-  2.0}^{+  2.0}$ & FERMI & \cite{Poolakkil2021} \\ 
GRB 140508A  & 1.03 & $ 22.5_{-  1.5}^{+  1.5}$ & KONUS & \cite{Tsvetkova2017} \\ 
GRB 140512A  & 0.725 & $ 10.0_{-  0.2}^{+  0.2}$ & FERMI & \cite{Poolakkil2021} \\ 
GRB 140614A  & 4.233 & $ 17.5_{-  6.4}^{+  4.6}$ & KONUS & \cite{Tsvetkova2021} \\ 
GRB 140703A  & 3.14 & $ 24.4_{-  1.1}^{+  1.1}$ & FERMI & \cite{Poolakkil2021} \\ 
GRB 141028A  & 2.33 & $ 67.5_{-  0.9}^{+  0.9}$ & FERMI & \cite{Poolakkil2021} \\ 
GRB 141109A  & 2.993 & $ 39.5_{-  3.2}^{+  3.8}$ & KONUS & \cite{Tsvetkova2021} \\ 
GRB 150206A  & 2.09 & $ 61.9_{-  4.5}^{+  4.5}$ & KONUS & \cite{Tsvetkova2017} \\ 
GRB 150314A  & 1.76 & $ 76.8_{-  2.1}^{+  2.1}$ & KONUS & \cite{Tsvetkova2017} \\ 
GRB 150403A  & 2.06 & $116.4_{-  6.2}^{+  6.2}$ & KONUS & \cite{Tsvetkova2017} \\ 
GRB 150413A  & 3.139 & $ 40.8_{-  7.5}^{+  7.2}$ & KONUS & \cite{Tsvetkova2021} \\ 
GRB 150821A  & 0.76 & $ 15.5_{-  1.2}^{+  1.2}$ & KONUS & \cite{Tsvetkova2017} \\ 
GRB 151021A  & 2.33 & $112.7_{-  9.4}^{+  9.4}$ & KONUS & \cite{Tsvetkova2017} \\ 
GRB 151111A  & 3.5 & $ 13.8_{-  1.6}^{+  1.8}$ & KONUS & \cite{Tsvetkova2021} \\ 
GRB 160131A  & 0.97 & $ 87.0_{-  6.6}^{+  6.6}$ & KONUS & \cite{Tsvetkova2017} \\ 
GRB 160227A  & 2.38 & $ 25.3_{-  5.8}^{+  6.3}$ & KONUS & \cite{Tsvetkova2021} \\ 
GRB 160327A  & 4.99 & $ 20.8_{-  4.7}^{+  4.9}$ & KONUS & \cite{Tsvetkova2021} \\ 
GRB 160509A  & 1.17 & $113.0_{- 10.0}^{+ 10.0}$ & KONUS & \cite{Tsvetkova2017} \\ 
GRB 160623A  & 0.37 & $ 25.3_{-  0.3}^{+  0.3}$ & KONUS & \cite{Tsvetkova2017} \\ 
GRB 160625B  & 1.41 & $510.1_{-  6.2}^{+  6.2}$ & KONUS & \cite{Tsvetkova2017} \\ 
GRB 160629A  & 3.33 & $ 38.9_{-  1.4}^{+  1.4}$ & KONUS & \cite{Tsvetkova2017} \\ 
GRB 161014A  & 2.823 & $ 10.7_{-  0.5}^{+  0.5}$ & FERMI & \cite{Poolakkil2021} \\ 
GRB 161017A  & 2.013 & $ 16.5_{-  3.2}^{+  4.2}$ & KONUS & \cite{Tsvetkova2021} \\ 
GRB 161023A  & 2.7106 & $ 69.5_{-  9.9}^{+  9.9}$ & KW1 & \cite{deUgarte2018,Frederiks2016} \\ 
GRB 161117A  & 1.549 & $ 23.6_{-  0.3}^{+  0.3}$ & FERMI & \cite{Poolakkil2021} \\ 
GRB 170202A  & 3.645 & $ 22.7_{-  3.1}^{+  3.7}$ & KONUS & \cite{Tsvetkova2021} \\ 
GRB 170214A  & 2.53 & $414.0_{-  2.3}^{+  2.3}$ & FERMI & \cite{Poolakkil2021} \\ 
GRB 170405A  & 3.51 & $252.0_{-  2.9}^{+  2.9}$ & FERMI & \cite{Poolakkil2021} \\ 
GRB 170705A  & 2.01 & $ 12.5_{-  0.4}^{+  0.4}$ & FERMI & \cite{Poolakkil2021} \\ 
GRB 171010A  & 0.3285 & $ 22.1_{-  0.1}^{+  0.1}$ & FERMI & \cite{Poolakkil2021} \\ 
GRB 180314A  & 1.445 & $ 10.4_{-  0.2}^{+  0.2}$ & FERMI & \cite{Poolakkil2021} \\ 
GRB 180325A  & 2.248 & $ 22.3_{-  3.5}^{+  3.6}$ & KW1 & \cite{DAvanzo2018a,Frederiks2018a} \\ 
GRB 180624A  & 2.855 & $ 19.4_{-  3.6}^{+  4.7}$ & KONUS & \cite{Tsvetkova2021} \\ 
GRB 180720B  & 0.654 & $ 60.1_{-  2.8}^{+  2.9}$ & KW1 & \cite{Vreeswijk2018,Frederiks2018b} \\ 
GRB 180914B  & 1.096 & $336.9_{- 14.7}^{+ 14.6}$ & KW1 & \cite{Davanzo2018b,Frederiks2018c} \\ 
GRB 181020A  & 2.938 & $ 69.6_{-  9.0}^{+  9.7}$ & KW1 & \cite{Fynbo2018,Tsvetkova2018} \\ 
GRB 181110A  & 1.505 & $ 20.7_{-  2.6}^{+  3.7}$ & KONUS & \cite{Tsvetkova2021} \\ 
GRB 181201A  & 0.45 & $ 12.8_{-  0.4}^{+  0.4}$ & KW1 & \cite{Izzo2018,Svinkin2018} \\ 
GRB 190106A  & 1.859 & $ 11.2_{-  1.7}^{+  2.5}$ & KW1 & \cite{Schady2019,Tsvetkova2019} \\ 
GRB 190114C  & 0.4245 & $ 21.5_{-  0.3}^{+  0.3}$ & KW1 & \cite{CastroTirado2019,Ursi2020} \\ 
GRB 190530A  & 0.9386 & $134.4_{-  3.6}^{+  3.7}$ & KW1 & \cite{Gupta2022,Frederiks2019a} \\ 
GRB 191004B  & 3.5055 & $ 11.4_{-  1.1}^{+  1.2}$ & KW1 & \cite{Vielfaure2020a,Svinkin2019} \\ 
GRB 191221B  & 1.148 & $ 34.8_{-  3.5}^{+  3.5}$ & KW1 & \cite{Vielfaure2019,Frederiks2019b} \\ 
GRB 200524A  & 1.256 & $ 13.7_{-  2.5}^{+  2.9}$ & KW1 & \cite{Yao2021,Svinkin2020a} \\ 
GRB 200613A  & 1.2255 & $ 20.3_{-  0.3}^{+  0.2}$ & FER & \cite{deUgarte2021a,Bissaldi2020} \\ 
GRB 200829A  & 1.29 & $126.0_{-  3.0}^{+  2.9}$ & KW1 & \cite{Pankov2023,Ridnaia2020} \\ 
GRB 201103B  & 1.105 & $ 17.0_{-  2.5}^{+  2.7}$ & KW1 & \cite{Xu2020,Svinkin2020b} \\ 
GRB 201216C  & 1.10 & $ 60.6_{-  3.3}^{+  3.3}$ & KW1 & \cite{Vielfaure2020b,Frederiks2020a} \\ 
GRB 201221A  & 5.70 & $ 25.1_{-  4.7}^{+ 10.9}$ & KW1 & \cite{Malesani2020,Frederiks2020b} \\ 
GRB 210610B  & 1.1345 & $ 46.2_{-  3.5}^{+  3.6}$ & KW1 & \cite{deUgarte2021b,Frederiks2021a} \\ 
GRB 210619B  & 1.937 & $385.2_{- 10.9}^{+ 10.9}$ & KW1 & \cite{deUgarte2021c,Svinkin2021} \\ 
GRB 210702A  & 1.16 & $ 77.1_{-  6.2}^{+  6.1}$ & KW1 & \cite{Xu2021,Frederiks2021b} \\ 
GRB 210822A  & 1.736 & $ 85.3_{-  7.8}^{+  7.8}$ & KW1 & \cite{Zhu2021,Frederiks2021c} \\ 
GRB 210905A  & 6.318 & $127.0_{- 19.0}^{+ 20.0}$ & KW1 & \cite{Tanvir2021,Rossi2022} \\ 
GRB 220101A  & 4.618 & $324.0_{- 20.0}^{+ 21.0}$ & KW1 & \cite{Fynbo2022,Tsvetkova2022a,Tsvetkova2022b} \\ 
GRB 220107A  & 1.246 & $ 23.9_{-  1.3}^{+  1.5}$ & KW1 & \cite{CastroTirado2022,Ridnaia2022} \\ 
GRB 220117A  & 4.961 & $ 15.3_{-  3.8}^{+  3.8}$ & KW1 & \cite{Palmerio2022,Frederiks2022a} \\ 
GRB 220527A  & 0.857 & $ 12.8_{-  0.7}^{+  0.6}$ & KW1 & \cite{Xu2022,Lysenko2022} \\ 
GRB 220627A  & 3.084 & $244.0_{- 27.0}^{+ 28.0}$ & KW1 & \cite{Izzo2022,Frederiks2022b} \\ 
GRB 230204B  & 2.142 & $227.0_{-  2.0}^{+  3.0}$ & FER & \cite{Saccardi2023,Poolakkil2023} \\ 
GRB 230812B  & 0.360 & $ 11.5_{-  0.3}^{+  0.2}$ & KW1 & \cite{deUgarte2023,Frederiks2023a} \\ 
GRB 231215A  & 2.305 & $114.0_{- 10.0}^{+ 11.0}$ & KW1 & \cite{Thoene2023,Frederiks2023b} \\ 
\noalign{\smallskip}\hline\noalign{\smallskip}
GRB 221009A  & 0.151 & $280_{-50}^{+80}$ & ALPHA & \cite{Ripa2023} \\ 
             &        & $740_{-10}^{+10}$ & SIRI & \cite{Mitchell2024} \\ 
             &        & $1500_{-16}^{+20}$ & GECAM-C & \cite{An2023} \\ 
             &        & $1010_{-7}^{+7}$ & FERMI & \cite{Lesage2023} \\ 
             &        & $1200_{-100}^{+100}$ & KONUS & \cite{Frederiks2023b} \\
             \noalign{\smallskip}
\enddata
\label{tab:tab1}
\end{deluxetable*}


\bibliography{Eisomax2023v3}{}
\bibliographystyle{aasjournal}



\end{document}